\documentstyle[11pt,paspconf,epsf]{article}
\def\edcomment#1{\iffalse\marginpar{\raggedright\sl#1\/}\else\relax\fi}
\reversemarginpar

\begin{document}
\title{CO Emission in Powerful Radio Galaxies at Low and High Redshift}
\author{A. S. Evans}
\affil{California Institute of Technology 105-24, Pasadena, CA 91125 U.S.A.}

\begin{abstract}
CO observations of $0<z<4$ radio galaxies are presented.
The $L_{\rm CO}$ upper limits for the high-redshift powerful radio galaxies 
(HzPRGs: $z > 1$, $P_{408{\rm MHz}} > 10^{27}$ Watts Hz$^{-1}$) 
are consistent
with no evolution in 
molecular gas (H$_2$) mass with redshift and/or radio power.
Of the low-redshift powerful radio galaxies (LzPRGs: $z<0.2$,
$P_{408{\rm MHz}} > 10^{23.5}$ W Hz$^{-1}$) 
observed, only one F-R II 
galaxy has been detected in CO, whereas $\sim$ 50\% of the
radio-compact (FC) and F-R I 
galaxies have been detected.
The CO and imaging data to date imply that either F-R II galaxies result
from mergers of gas-poor galaxies relative to 
FC/F-R I galaxies, or that some FC/F-R I galaxies evolve
into F-R II galaxies. In either scenario, 
the radio activity in powerful radio
galaxies is most likely 
triggered by a merger event, and
H$_2$ may be the fuel for the active galactic nucleus (AGN) 
in the early stages of the
merger.
\end{abstract}

\section{Introduction}

Advances in millimeter receiver technology and observing techniques, 
as well as the 
availability of moderate-size millimeter telescopes and arrays,
have 
made recent CO observations of distant extragalactic sources possible.
During the mid-1980s, CO surveys of galaxies
selected by their {\it IRAS} fluxes and colors resulted in 
CO detections
of a host of $z < 0.2$ Seyfert, HII region-like, and LINER galaxies,
as well as several QSOs and radio galaxies (e.g., Sanders, Scoville,
\& Soifer 1991). 
Their inferred H$_2$ masses are $1\times10^9 - 6\times10^{10}$
M$_\odot$. For galaxies believed to have AGNs, H$_2$
provides a possible source of fuel for nuclear activity, as well as
for vigorous, circumnuclear star formation.
The detection of CO emission 
in the $z=2.3$ {\it IRAS} galaxy FSC 10214+4724 (Brown \& Vanden Bout
1991, 1992) revolutionized the field,
both because of its cosmological implications
and because it proved that it was possible, with modern technology, to
detect star-forming gas out to redshifts of a few.
Since then,
much effort has gone into searching for CO in a wide 
variety of high-redshift 
objects.

At the time FSC 10214+4724 was detected in CO, 
the only sample of galaxies known at comparable redshifts were radio galaxies.
HzPRGs were well suited as initial targets for CO surveys, primarily because
{\it (i)} several LzPRGs had been detected in
CO (Phillips et al. 1987; Mirabel, Sanders, \& Kaz\`{e}s 1989;
Mazzarella et al. 1993),
{\it (ii)} 
radio galaxies are observed to span a wide redshift range ($0 < z <4$)
and {\it (iii)}  
their redshifts are known to the accuracy required by the 512 MHz bandwidths
of millimeter detectors.
Although the number of LzPRGs 
observed in CO at that time was small,
further evidence that radio galaxies as a class
may be rich in H$_2$ came from the
observations that
a significant fraction (20--30\%) of LzPRGs were known
to be infrared luminous (Golombek, Miley, \& Neugebauer 1988). 
Some have infrared excesses consistent with
thermal emission from dust, and many LzPRGs 
possess morphological peculiarities commonly associated with 
collisions/mergers of galaxy pairs (Heckman et al. 1986). 
Multi-component structures and/or nearby companions are also observed in
$z \sim 1$ HzPRGs (Spinrad \& Djorgovski 1984;
Lilly \& Longair 1984; Djorgovski et al. 1987), 
but the {\it IRAS} flux upper
limits yield very little information on the shape and luminosity of the
far-infrared emission in these galaxies (Evans et al. 1996).
The blue colors of many HzPRGs are often cited as additional, though 
less compelling, evidence that they may be H$_2$ rich; the
colors may also be due to scattered AGN light. 

In this article, the results of the CO observations of
HzPRGs (e.g., Evans et al. 1996; Downes et al. 1996) 
will first be summarized, followed by a discussion of
a complete 60 and 100 $\micron$ flux-limited sample of
LzPRGs (e.g. Evans et al. 1998a). 
The latter discussion will aim at clarifying the results 
obtained from the high-redshift observations. 
With the exception of Figure 1,
$H_0 = 75 $ km s$^{-1}$ Mpc$^{-1}$, $q_0 = 0.5$, and $\alpha$ =
4 M$_\odot$ (K km s$^{-1}$ pc$^2)^{-1}$ are assumed throughout.

\section{The High-Redshift Radio Galaxy Survey}

\begin{figure}
\plotfiddle{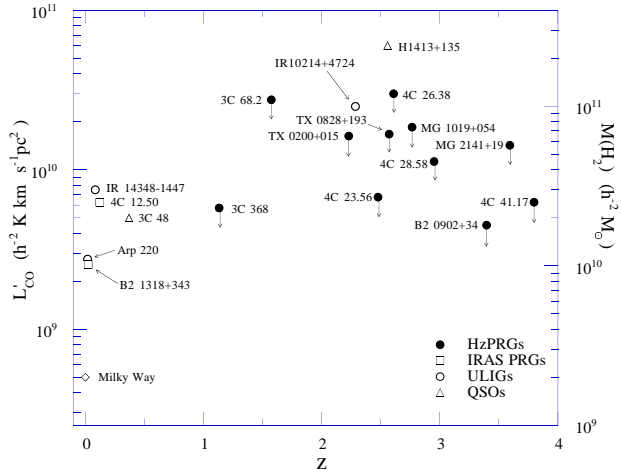}{2.2 in}{0}{50}{50}{-200}{-60}
\caption{$L'_{\rm CO}$ and $M({\rm H}_2)$ vs. $z$ (Evans et al. 1996).
$\downarrow$ denotes 3$\sigma$ upper limits. $H_0 = 100h$ km s$^{-1}$
Mpc$^{-1}$ and $q_0 = 0.5$ is assumed.}
\end{figure}

The primary goals of any CO survey of high-redshift galaxies 
are to estimate the H$_2$ masses 
and to determine if 
the H$_2$ content evolves as a function of redshift.
If the radio power of radio galaxies is correlated with the
amount of H$_2$,
then HzPRGs, which are
preselected by their high radio luminosity, may have rich
circumnuclear ISMs that serve as fuel for the AGN. 
Thus, a sample of 15 HzPRGs 
spanning the redshift range $1<z<4$ were selected. 
For each galaxy, observations were done by tuning to $J > 1$ rotational
CO transitions redshifted to frequencies accessible by the 3mm
receivers on the NRAO 12m and IRAM 30m Telescopes.
Eleven HzPRGs 
were eventually observed at one or both telescopes, with integration
times of up to 10 hours at the 30m Telescope, and up to 50 hours at
the 12m Telescope.

None of the eleven galaxies observed were detected
(Figure 1). However, assuming
a line width of $v_{\rm FWHM} \sim$ 250 km s$^{-1}$ (the average 
$v_{\rm FWHM}$ observed for
a complete sample of ultraluminous infrared galaxies, ULIGs: Sanders, Scoville,
\& Soifer 1991), the 3$\sigma$ upper limits for H$_2$ mass are in the
range 3.5--20$\times10^{10}$ M$_{\odot}$ ($H_0 = 75$ km s$^{-1}$ 
Mpc$^{-1}$). Further,
the four HzPRGs with the most sensitive CO upper limits
would appear to have
less H$_2$ than the most H$_2$-rich galaxies in the local
universe. This would seem to imply that the H$_2$ content
of radio galaxies does not evolve as a function of redshift
or radio power. 

\begin{figure}
\plotfiddle{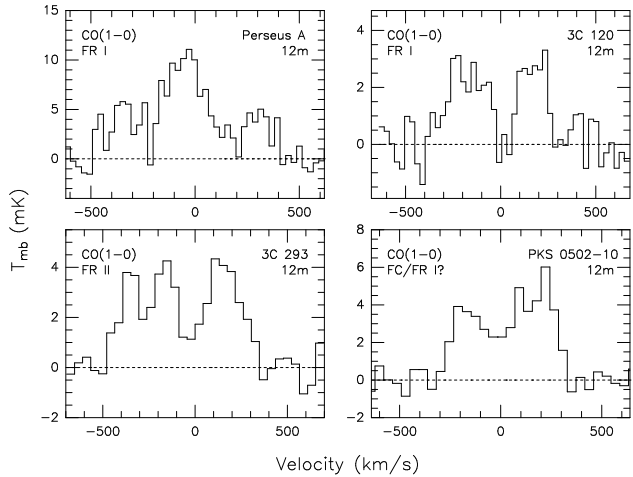}{2.2 in}{0}{50}{50}{-170}{-190}
\caption{Examples of CO emission lines in LzPRGs 
(Evans et al. 1998a).}
\end{figure}

Is it the case that, while radio-weak ULIGs show
evidence of evolution of H$_2$ mass with redshift, 
radio-loud infrared galaxies, such as radio
galaxies, do not? 
Indeed, the recent tentative detection of CO in 53W002
($\sim 7\times10^{10}$ M$_\odot$ of H$_2$: Scoville et al. 1997), a
relatively radio-faint HzPRG in comparison with the 11 HzPRGs discussed above,
may be evidence of an anticorrelation between H$_2$ mass and radio
power.
In actuality, there is very little evidence for 
evolution of H$_2$ mass
in any class of galaxies as a function of redshift. 
IRAS 10214+4724
is now known to be gravitationally lensed
(e.g. Eisenhardt et al. 1996); its inferred intrinsic
H$_2$ mass is $\sim 10^{10}$ M$_{\odot}$,
a fairly average value for an ULIG. In 1994, a $z=2.6$
gravitationally-lensed QSO (H1413+117) was detected (Barvainis et al. 1994),
with an intrinsic luminosity also believed to be average for ULIGs 
at low redshift. Two other ULIGs
(IR 09104+4109 at $z=0.4$ and IR 15307+3252 at $z=0.9$) and one high-redshift
{\it IRAS}-luminous QSO (PG 1634+706 at $z=1.3$) have been searched for CO, but
the 3$\sigma$ upper limits of $L'_{\rm CO}$ are less than that of
the most gas-rich
galaxies observed locally (Evans et al. 1998b). 
Finally, the $z=4.7$ 
QSO BR 1202-07 has recently been detected
and confirmed, and its inferred H$_2$ mass is 
$\sim 1\times10^{11}$ M$_\odot$
(Omont et al. 1996; Ohta et al. 1996). This may be the first indication
of evolution in H$_2$ mass with redshift, however, the H$_2$ mass of 
BR 1202-07 is only twice 
that observed in the most H$_2$-rich, local galaxies. There
is also evidence for shear due to gravitatiional 
lensing in the vicinity of the QSO, which
may indicate that its intrinsic $L'_{\rm CO}$ is much lower than observed
(Omont et al. 1996). 

\begin{figure}
\plotfiddle{}{4.9 in}{0}{100}{100}{-300}{-290}
\caption{The F-R II galaxy 3C 293 (Evans et al.
1999). The scale is 860 pc/$\arcsec$. The radio continuum map is from
Leahy, Pooley, \& Riley (1986).}
\end{figure}

If, as is commonly believed, radio galaxies
are the progenitors of some giant elliptical galaxies,
there are four possible reasons why CO emission has not been 
unambiguously detected
in any HzPRGs, and why evolution of H$_2$ mass with redshift is not seen in
other classes of galaxies:
First, the redshift at which the bulk of stars in present day
galaxies formed may be greater than four. Thus, most of the H$_2$ 
may have been consumed by starbursts and the AGN
by the redshift most of these galaxies are located.
Second, because merger events must have been more common in the past,
most of the H$_2$ that eventually formed the bulk of the stars
in giant elliptical galaxies
may be in many smaller, star-forming galaxies. These
individual galaxies may consume most of their H$_2$ before
merging.
Third, the duration that
galaxies can contain in excess of $4-6 \times 10^{10}$ M$_\odot$
of H$_2$
may be very brief. Thus, the probability of observing such a gas-rich
galaxy may be exceedingly low.
Fourth, the universe may not have closure density, thus the scale
of the universe may be larger
than assumed for the above H$_2$ mass calculations.
Even given this possibility, the H$_2$ mass
upper limits of the four best studied
HzPRGs are substantially
less than the total stellar mass of a giant elliptical
galaxy.

\section{The Low-Redshift Radio Galaxy Survey}

From the high-redshift observations alone, it was not clear what role
H$_2$ may play in the radio activity seen in these galaxies. In addition,
the number of radio galaxies observed prior to the HzPRG CO survey
was very small, thus
it was not clear if inaccurate generalizations based on observations
of a few gas-rich LzPRGs 
were being made for radio
galaxies as a whole. Of the several LzPRGs 
that had been observed, none of those
detected had been F-R II (i.e., edge-brightened
radio morphology, typically observed to have
$P_{\rm 408 MHz} > 10^{25.3}$ W Hz$^{-1}$: Fanaroff \& Riley 1974)
galaxies like the HzPRGs - all had been
FC or F-R I (i.e., edge-darkened radio morphologies,
typically with
$P_{\rm 408 MHz} < 10^{25.3}$ W Hz$^{-1}$) galaxies.

Over the last few years,
considerable time has been devoted to 
investigating the H$_2$ masses and morphologies, as well as
the 2.2$\micron$ morphologies of a complete
sample of 35 LzPRGs with {\it IRAS} fluxes at 60$\mu$m
and/or 100$\mu$m $> 0.30$ Jy and $\delta > -40^{\rm o}$
(Evans et al. 1998a). Examples of some
of the CO emission lines detected are shown in Figure 2. 
The single-dish observations are being obtained using the NRAO 12m Telescope,
with follow-up interferometry of detected sources obtained using the Owens
Valleys Millimeter Array.
The 2.2$\micron$ imaging, obtained 
to look for evidence of isophotal irregularities, 
imbedded multiple nuclei and bridges hidden
by dust at optical wavelengths, was done using the UH 2.2m Telescope. 
The results
from this survey are that only 40\% of the F-R II
galaxies show
evidence of disturbed, high surface brightness features and multiple
nuclei or close companions and that, aside from a CO detection
in 3C 293 ($z \sim 0.04$: Evans et al. 1999, 
Figure 2 and 3), none of the F-R
II sources have confirmed CO detections. Cygnus A, the
protypical F-R II radio galaxy, is among the galaxies 
that were not detected; although it has an $L'_{\rm IR} \sim 4.5\times10^{11}$
L$_\odot$, it has less than one-third the H$_2$ mass of the
Milky Way. The 2.2 $\micron$ image (not shown) 
supports this result - the host galaxy looks 
like an undisturbed elliptical galaxy. Further, Cygnus A and many
other LzPRGs have infrared colors suggestive of warm
dust temperatures, indicating that
most of the infrared luminosity may be coming from a small amount
of dust heated by the AGN 
(i.e., $L_{\rm IR} \propto M_{dust} T^5_{dust}$).   
In contrast, 86\% of the
FC and F-R I
sources show evidence of disturbed, high
surface brightness features and multiple nuclei or close companions,
and 50\% have been detected in CO.  Such a dramatic contrast could
simply suggest
that the progenitors of
F-R II galaxies are gas-poor relative to the progenitors of
FC and F-R I sources, or that some FC and F-R I
sources evolve into F-R II sources after the H$_2$ has been consumed.
The fundamental difference between F-R I and F-R II sources may
be the efficiency to which the jets propagate through the ISM of the host
galaxy. 
Given this, the lack of CO detections in HzPRGs is not surprising.

If the apparent deficiency in the H$_2$ masses of F-R II relative to 
FC/F-R I galaxies is due to time evolution, 
galaxies such as 3C 293 may be transitional
sources. The galaxy 3C 293 (Figure 3)  
has a moderately high infrared luminosity, a distorted
disk-like infrared morphology with a companion, and several times the
H$_2$ mass of the Milky Way. 
The CO emission is extended along the axis of
the 2.2 $\micron$ disk and perpendicular to the radio jet, possibly
because the jets have escaped through an axis unobscured by the putative
nuclear torus. 
The CO emission line (Figure 2), has an absorption feature approximately
150 km s$^{-1}$ wide. If the width of the line is due solely to rotation,
the absorption may occur from gas $\sim 50$ pc from the AGN, 
approximately the radius of the dust torus observed in HST images of the radio
galaxy NGC 4261 (Jaffe et al. 1996).

\acknowledgements

It is a pleasure to thank my collaborators
D. Sanders, J. Mazzarella, P. Solomon,
D. Downes, S. Simon, J. Surace, F. Mirabel, and C. Kramer.
I also am grateful to the staffs of the NRAO 12m, IRAM 30m,
and UH 2.2m Telescopes,
and the staff and postdocs of the 
Owens Valley Millimeter Array for their assistance during
and after the observations presented here.

\end{document}